# Transient Detour and Cooperative Oxygen Exchange in the Polarization Switching of Ferroelectric $Hf_{0.5}Zr_{0.5}O_2$

[1]Ryotaro Sahashi, [1]Po-Yen Chen, and [1,2]Teruyasu Mizoguchi (corresponding author)
Email: teru@iis.u-tokyo.ac.jp
[1]Department of Materials Engineering, The University of Tokyo, Tokyo, Japan
[2]Institute of Industrial Science, The University of Tokyo, Tokyo, Japan

**Abstract**
Hafnium zirconium oxide (HZO) has attracted significant attention as a core material for next-generation non-volatile memories due to its excellent ferroelectricity in the ultra-thin film regime and its CMOS process compatibility. However, the exploration of its polarization switching mechanism has predominantly relied on static energy barrier analyses, leaving the transient bond formation and cooperative dynamic mechanisms under actual electric field driving unresolved. In this study, we performed Electric-Field-Induced MD simulations on a defect-free ideal HZO lattice using a fine-tuned machine learning force field (MACEField). As a result, we successfully reproduced the *P-E* hysteresis loop dynamically and demonstrated that the polarization switching in HZO is driven not by conventional simple displacement models (S:N/S:T models), but by the dynamic mutual exchange of 3-coordinated oxygen (O3c) and 4-coordinated oxygen (O4c). Analysis of the oxygen atom displacement trajectories revealed that this pathway is accompanied by a unique "detour" behavior originating from transient cation-oxygen bond formation. Furthermore, we identified an "internal self-compensation mechanism" in which the local volumetric expansion and contraction accompanying the coordination number changes are effectively offset within the cell. These findings provide, from a dynamic perspective, a microscopic physical origin of for HZO's exceptional ability to sustain stable polarization switching without macroscopic strain, a property that has long distinguished HZO from conventional perovskite ferroelectrics yet lacked atomistic explanation. These findings suggest that preserving the integrity of cooperative O3c/O4c exchange pathways¸rather than minimizing individual atomic displacements, is the key design principle for endurance and scalability in next-generation ferroelectric memories.

## 1. Introduction

Hafnium oxide ($HfO_2$)-based ferroelectrics, [1] particularly hafnium zirconium oxide (HZO) in which zirconium forms a solid solution, have attracted tremendous attention as core materials for next-generation ultra-high-density and low-power non-volatile memories. The primary reasons HZO is considered especially promising among various $HfO_2$-based ferroelectrics lie in its excellent process compatibility and the stability of its properties in ultra-thin films. [2,3] Compared to $HfO_2$ doped with elements such as Si or Al, HZO exhibits a lower crystallization temperature for the ferroelectric phase [4–6], offering high compatibility with the back-end-of-line (BEOL) processes in semiconductor manufacturing. Furthermore, while the ferroelectric properties of conventional perovskites are strongly affected by depolarization fields and interface effects upon thickness scaling, [7,8] $HfO_2$-based ferroelectrics can retain switchable polarization in films with thicknesses below 10 nm and even in the sub-5-nm regime. [9,10] $HfO_2$-based ferroelectrics can maintain switchable polarization under mechanically constrained thin-film geometries, although their phase stability and polarization are strongly influenced by interfacial stress and epitaxial strain. [11–13] These highly advantageous practical characteristics make HZO the leading candidate for actual device integration.

Understanding the origin of this unique ferroelectricity in HZO and the microscopic mechanism of polarization switching under an applied electric field is indispensable for further improving material performance and ensuring device reliability. Polarization switching in the ferroelectric phase (orthorhombic Pca21 phase) of HZO is induced by the displacement of oxygen atoms within the crystal lattice. Until now, the dominant discussion regarding its switching pathways has focused on simple displacement models that do not involve a change in the coordination numbers of oxygen atoms (3-coordinated: O3c, 4-coordinated: O4c), such as the Shift-Through (S:T) model, where oxygen atoms cross the cation plane, and the Shift-Inside (S:N) model, where they do not. [14,15] Recently, however, the existence of complex pathways involving coordination number changes (exchanges) has been theoretically proposed by X. Mu et al. [16], suggesting that polarization switching may encompass more dynamic and complex mechanisms.

Nevertheless, the exploration of these polarization switching pathways has traditionally relied heavily on "static" energy barrier analyses, such as the Nudged Elastic Band (NEB) method based on density functional theory (DFT). [17,18] While static analysis is effective for identifying the minimum energy path connecting the initial and final states, it has fundamental limitations in capturing the effects of transient bond formation occurring under electric field driving, as well as the "dynamic" processes where multiple oxygen atoms operate cooperatively. To overcome the limitations of static approaches, ab initio molecular dynamics (AIMD) simulations have been employed to explore the intrinsic ferroelectricity and atomic dynamics of $HfO_2$. [19] However, classical AIMD remains constrained in spatial and temporal scales, and directly modeling large-scale response under explicit electric fields remains challenging. Recently, to resolve this issue, there has been a report applying Electric-Field-Induced Molecular Dynamics (MD) simulations using machine learning force fields (MLFFs) to $HfO_2$, [20] demonstrating the utility of large-scale, long-time dynamic analyses. Building on this trend, we conceived the idea of elucidating the dynamic polarization switching mechanism specifically in HZO, which excels in ultra-thin film stability. To perform the Electric-Field-Induced MD in this study, we adopted MACEField [21], which already has an established track record in polarization switching studies of BTO and natively supports the integration of electric field and charge interactions. [22]

In this paper, we applied fine-tuning tailored for HZO to this MACEField model and conducted Electric-Field-Induced MD simulations using a defect-free ideal lattice. This method enables the direct observation of the dynamical polarization-switching pathway, providing a clearer understanding of the switching mechanism in $HfO_2$-based ferroelectric materials.

## 2. Methods

### 2.1. Dataset Construction and First-Principles Calculations for $Hf_{0.5}Zr_{0.5}O_2$

To develop a MLFF specialized for the HZO system, a comprehensive structural database based on first-principles calculations was constructed from scratch. The reference electronic structure calculations were performed using the Vienna Ab initio Simulation Package (VASP) with the PBEsol [23] exchange-correlation functional.

First, on-the-fly learning [24] via heating MD simulations was conducted within a temperature range of 1 to 1000 K for the supercell structures of the four primary crystalline phases of HZO: $Pca2_1$ (96 atoms), $P2_1/c$ (96 atoms), $P4_2/nmc$ (108 atoms), and Pbca (96 atoms), utilizing a Gamma-centered 3×3 ×3 K-point mesh. To accurately represent the disordered solid solution of Hf and Zr within these phases, the atomic arrangements in all supercells were constructed using the SQS method. Through this process, a total of 2,257 structures were extracted, comprising 635 structures for the $Pca2_1$ phase, 562 for $P2_1/c$, 513 for $P4_2/nmc$, and 547 for Pbca.

For all extracted structures, density functional perturbation theory (DFPT) [25,26] calculations were performed using VASP to obtain the training dataset. The cut-off energy (ENCUT) was set to 800 eV, and the electronic convergence criterion (EDIFF) was $10^{-5}$ eV, with a Gamma-centered 2×2×2 k-point mesh. The collected dataset of 2,257 structures was initially partitioned randomly into a training/validation set and a test set in an 8:2 ratio. The training/validation set was further divided into a training set and a validation set in a 9:1 ratio to construct and optimize the model. The total energy, atomic forces, and Born effective charges (BECs) were extracted from the dataset as the learning targets. The remaining 20% of the structures reserved for the test set were utilized to evaluate the final predictive accuracy of the model, and the resulting mean absolute errors (MAEs) for the energy, forces, and BECs are compiled in the Supporting Information (SI).

### 2.2. Development of the Machine Learning Force Field Using MACEField

To predict the total energy, atomic forces, and electric-field responses (BECs) within a unified framework, a machine learning model was fine-tuned using the MACEField architecture. The large-scale foundation model MACE-MP-0a large [27] was adopted as the starting configuration and fine-tuned on the partitioned 2,257-structure DFPT dataset.

The training was conducted with a cut-off radius ($r_{max}$) of 6.0 Å, a batch size of 4, and the AMSGrad optimizer. The universal_field loss function was employed, with loss weights assigned as 100.0 for BECs, 10.0 for energy, and 10.0 for forces. To stabilize the training process, an exponential moving average (EMA) of the model weights was applied with a decay rate of 0.99, and stochastic weight averaging (SWA) was introduced after 500 epochs. To prevent overfitting, an early stopping criterion with a patience of 20 epochs based on the validation loss was implemented.

### 2.3. Quantitative Evaluation of the Degree of Orientation for Oxygen Displacement

To quantitatively analyze the oxygen atom dynamics during the polarization switching process, the "Degree of Orientation" $P_{<uvw>}$ was introduced to measure how well the displacement vectors of individual oxygen atoms align along a specific direction. [28] Here, the origin of displacement (the center of calculation) for a given oxygen atom was consistently defined as the center of mass of its four surrounding cations when the oxygen atom resides in its 4-coordinated (O4c) state.

Let $<uvw>$ be the direction vector of the oxygen displacement from this center of mass in the MD trajectory, and let $\alpha_{<uvw>}$ be the angle between this displacement vector and the reference axis of interest. The degree of orientation $P_{<uvw>}$ is calculated as follows:

$$P_{<uvw>} = < cos^2 \alpha_{<uvw>} > \quad (1)$$

A value approaching 1 indicates that the oxygen displacements are highly oriented along the targeted direction. Conversely, if the oxygen displacements are spatially completely random, the statistical theoretical values (random baselines) are $P_{<100>}$ = 0.7008 for the $<100>$ direction, $P_{<110>}$ = 0.8354 for the $<110>$ direction, and $P_{<111>}$ = 0.7577 for the $<111>$ direction. Using this index, the alignment and transient transition behaviors of the oxygen atom displacements during polarization switching were rigorously tracked.

### 2.4. Polarization Evaluation and Electric-Field-Induced Molecular Dynamics Simulations

To investigate the dynamic ferroelectric switching behavior under an external electric field, electric-field-induced MD simulations were performed using the fine-tuned MACEField model.

To determine the optimal system size that adequately captures the macroscopic ferroelectric response without size artifacts, the simulations were initially conducted across various $Pca2_1$ supercell sizes: 96, 768, and 2,592 atoms. Based on the convergence of the *P-E* hysteresis loops and the calculated coercive fields (**Figure S7**), the supercell containing 2,592 atoms was selected as the primary target for the detailed analyses. The simulations focused primarily on a $Pca2_1$ supercell containing 2,592 atoms.

An external electric field was continuously applied along the x-axis with a sweep rate of $5.0 \times 10^{-5}$ MV/(cm·fs). The spontaneous polarization $P$ at each step was calculated using the displacement $\Delta u_i$ of each atom $i$ along the polarization direction relative to a centrosymmetric, nonpolar reference configuration (the $P4_2/nmc$ phase was adopted in this study) and the dynamic BECs $Z_i^*$ predicted on-the-fly by the model, according to the following equation:

$$P = \frac{e}{V}\sum_{i} Z_i^* \, \Delta u_i \quad (2)$$

where $\boldsymbol{e}$ is the elementary charge and $V$ is the instantaneous cell volume. Through this approach, direct modeling of the *Polarization-Electric Field* (*P-E*) hysteresis loop incorporating the dynamic lattice responses and oxygen atom displacements under an electric field sweep was achieved.

## 3. Results and Discussion

To perform MD simulations under an applied electric field, interatomic forces predicted by the fine-tuned MACEField model were combined with electric-field-induced forces derived from BECs predicted by the same MACEField model.

The external electric-field-induced force acting on atom $i$, denoted as $F_{\text{ext}}$, was calculated at each MD step following the method proposed by Chen et al. [20,22,29] according to Equation (3):

$$F_{\text{ext}} = | \, e \, | \; \mathcal{E}_\beta Z_{i,\beta\alpha}^* \quad (3)$$

In Equation (1), $e$ is the elementary charge, $\mathcal{E}_\beta$ is the applied electric-field vector, and $Z_{i,\beta\alpha}^*$ is the Born Effective Charge (BEC) tensor predicted by the MACEField model.

The total force acting on each atom during time integration was given as the sum of the interatomic force calculated by the MACEField model and the external electric-field-induced force defined above.

All electric-field-coupled MD simulations were performed under the NPT ensemble. The applied electric field was swept cyclically following:

$$0 \rightarrow +10 \rightarrow -10 \rightarrow 0 \text{ MV cm}^{-1}.$$

### 3.1. Reliability of the Electric-Field-Induced MD Simulations

Prior to analyzing the polarization switching process, we verified the accuracy and reliability of the fine-tuned MACEField model. As shown in **Figure S1** of the Supplementary Information, the prediction accuracy of the MACEField model was evaluated against first-principles calculations using VASP [30–33]. The Mean Absolute Errors (MAE) for Force, Energy, and BEC were 12.1 meV/Å, 25.8 meV, and $7.33\times10^{-3}$ e, respectively, confirming an exceptionally high level of reproducibility. Furthermore, in the comparison of phonon calculations [34] for the major crystalline phases of HZO ($Pca2_1$, $P2_1/c$, $P4_2/nmc$, and Pbca) (**Figure S2**), the MAEs for each phase were 0.039, 0.037, 0.039, and 0.028 THz, respectively. This demonstrates that the model also possesses outstanding predictive accuracy from a lattice dynamics perspective.

In addition to these DFT-level verifications, our model successfully reproduces critical experimental measurements of the ferroelectric $Pca2_1$ phase. Specifically, the calculated $d_{111}$ interplanar spacing is 2.94 Å, showing remarkable quantitative agreement with the experimental value. [35,36] Moreover, the model properly captures macroscopic finite-temperature behaviors under applied electric fields; as shown in **Figure S3**, our simulations reproduce the experimental temperature-dependent trends of an increasing dielectric constant and a decreasing coercive field at elevated temperatures. [37,38]

Importantly, electric-field-induced MD simulations employing MLFFs have already been widely deployed and validated across a diverse range of ferroelectric and piezoelectric materials—such as $BaTiO_3$, $HfO_2$, and ScAlN—establishing this framework as a robust and cutting-edge methodology for capturing finite-temperature field responses. [20,22,29,39–41] The high predictive accuracy demonstrated across all our validation tests confirms that this powerful and proven methodology operates with exceptional reliability for HZO, providing a solid foundation for the subsequent polarization switching analyses.

**Figure 1**(a) shows the *P-E* hysteresis loop of HZO obtained from the electric-field-induced MD simulations using this highly accurate potential. The calculated coercive field ($E_c$) was 7.59 MV/cm, and the remnant polarization ($P_r$) was 48 μC/cm$^2$. The physical reason these values are larger than typical experimental values ($E_c$ ~ 1.00 MV/cm, $P_r$ ~ 24.5 μC/cm$^2$) [35] is that this simulation targets an ideal, perfect crystal structure containing no point defects, with the Hf and Zr atoms distributed as a random solid solution generated via the Special Quasirandom Structures (SQS) method [42]. In fact, H.B. Kim et al. reported that HZO thin films with extremely reduced oxygen vacancies exhibit exceptionally high values of $E_c$ = 2.01 MV/cm and $P_r$ = 36.5 μC/cm². [43] Additionally, DFT calculations by H. Lu et al. reported a $P_r$ of 55 μC/cm², further supporting the fact that the intrinsic polarization value in an ideal defect-free lattice is inherently larger than experimental values. [44]

As these findings and the comparisons in Table 1 indicate, the present model excellently reproduces the hysteresis loop—the dynamic behavior of ferroelectrics—and the calculated physical properties are highly reasonable, quantitatively and correctly reflecting the physical nature of a defect-free ideal lattice.

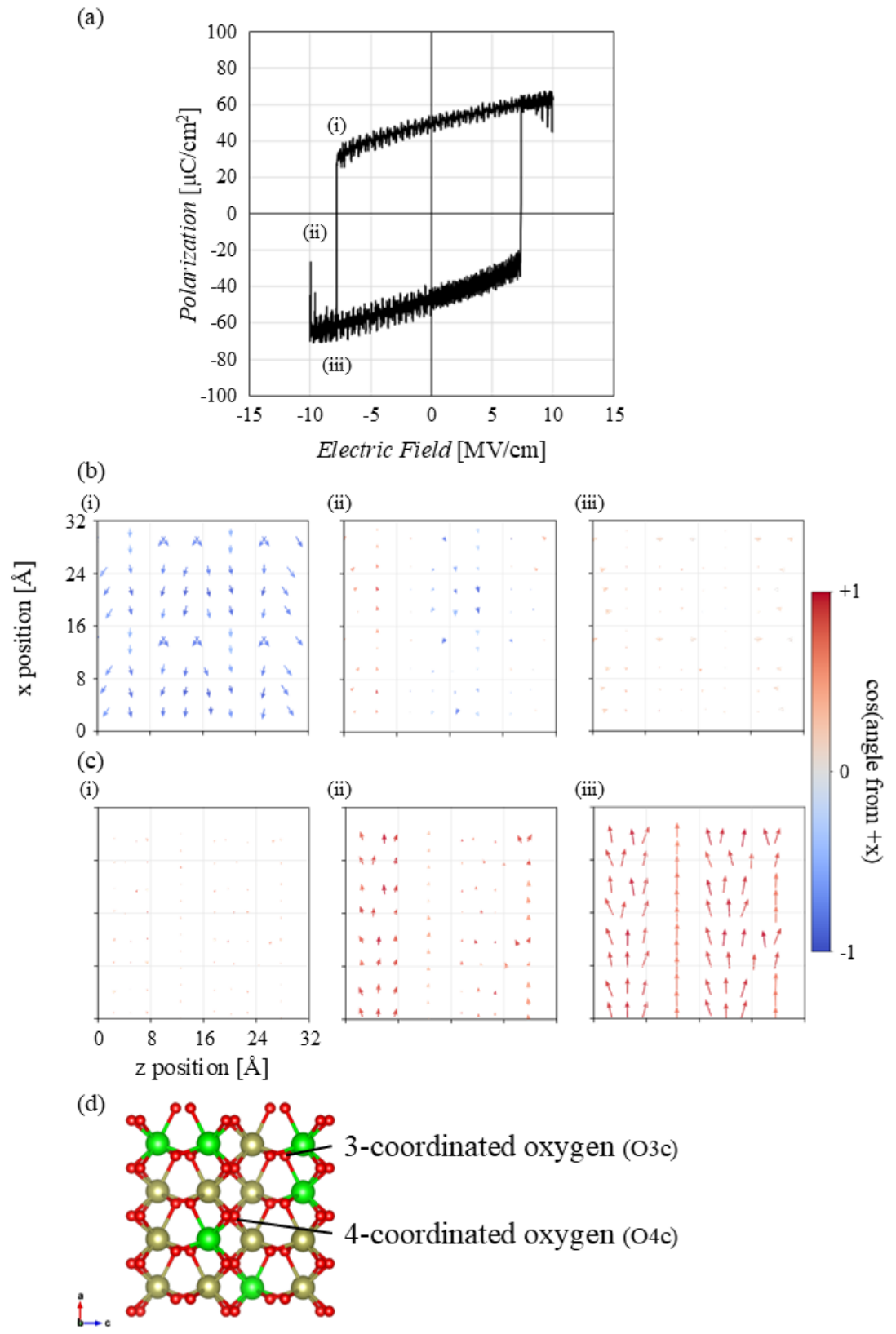

**Figure 1.** Polarization switching in HZO: (a) *P-E* hysteresis loop, and oxygen atom displacement vectors during the (b) O4c → O3c process and (c) O3c → O4c process, and (d) schematic diagram showing the local structures of the O3c and O4c states. In (b) and (c), the origin of each vector is the center of mass of the four surrounding cations to which the oxygen atom belongs. The arrows are color-coded with a gradient from 1 (red) to -1 (blue) based on the cosine of the angle from the +x-axis direction.

### 3.2. Exchange Behavior of O3c and O4c During Polarization Switching

In the ferroelectric phase ($Pca2_1$) of HZO, there are O3c and O4c. Until now, discussions on the polarization switching mechanism of HZO have been dominated by the S:N and S:T models, which do not account for coordination number changes accompanying oxygen atom movement. In contrast, while some previous studies discussed the theoretical possibility of pathways involving coordination changes, the specific dynamic pathways had not been definitively identified. The electric-field-driven MD simulations in this study directly and clearly revealed the dynamics in which the mutual exchanges of O3c → O4c and O4c → O3c occur cooperatively during the polarization switching process.

The specific displacement vectors of the oxygen atoms during the transition of the polarization from positive to negative are shown in Figure 1(b) (O4c → O3c) and Figure 1(c) (O3c → O4c). To accurately evaluate the geometric symmetry, the center of mass of the four surrounding cations in the O4c state was uniformly adopted as the coordinate origin for all displacement analyses. Furthermore, to quantitatively trace the detailed transitions of the spatial distribution and orientation of oxygen atoms accompanying this coordination change, we introduced a metric termed the "degree of orientation ($P$)". This metric measures how well the oxygen displacement vectors align with specific crystallographic directions. A value of $P$ = 1 indicates perfect alignment with the target direction. Conversely, a completely random spatial distribution of displacements yields geometric theoretical baselines of approximately 0.70, 0.84, and 0.76 for the *<100>*, *<110>*, and *<111>* directions, respectively. Based on this metric, we calculated the pole figures (**Figure 2**(a), (b)) and plotted the dynamic evolution of $P$ as well as its deviation from the random baselines (Figure 2(c): O4c → O3c, Figure 2(d): O3c → O4c).

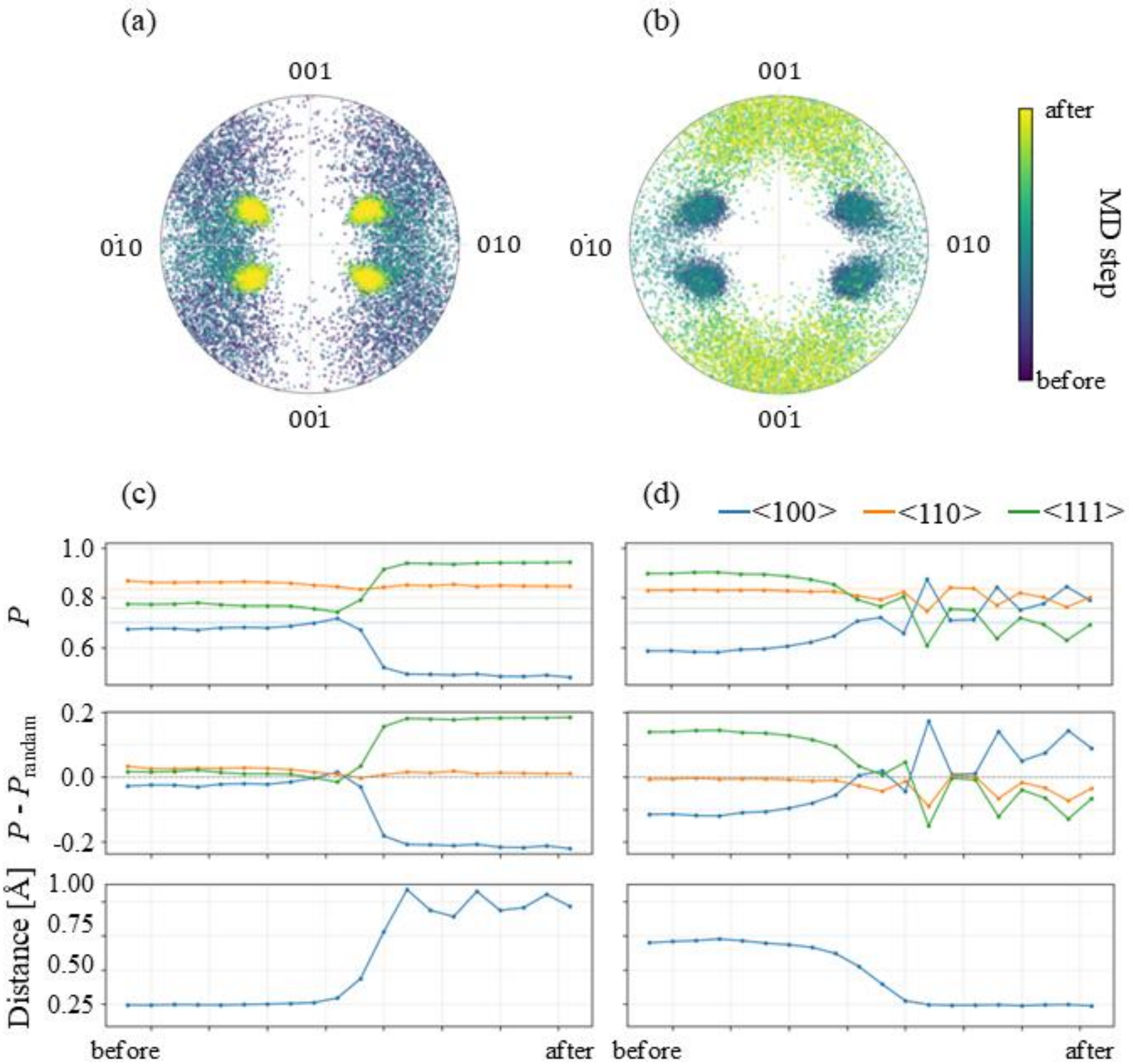

**Figure 2.** Displacement trajectories (pole figures) and dynamic transition of orientation for oxygen atoms during polarization switching. Pole figures of oxygen atom displacement trajectories viewed from the [100] direction during the (a) O4c → O3c process and (b) O3c → O4c process. For a total of 20 structures (10 structures each before and after the polarization switching point where the sign of the polarization value flips), the data are plotted with a gradient from purple to yellow as the polarization switching process progresses (from the before to after stages). Three-tier connected plots showing the degree of orientation $P$ (top), the degree of orientation minus the random baseline (middle), and the change in distance from the origin (bottom) during the (c) O4c → O3c process and (d) O3c → O4c process. Each direction is indicated by lines of <100> (blue), <110> (orange), and <111> (green).

The comprehensive analysis of the pole figures (Figure 2(a), (b)), the degree of orientation, and the distance from the origin (Figure 2(c), (d)) clearly reveals the geometric dynamics of this process and the distinct roles of each metric. Specifically, the pole figures (Figure 2(a), (b)) visually demonstrate the spatial symmetry and directionality of the trajectories. In parallel, the degree of orientation quantitatively traces the dynamic alignment transitions among specific crystallographic directions. Furthermore, the distance—measured from the center of mass of the four surrounding cations in the O4c state (Figure 2(c), (d))—directly captures the spatial shifts of the oxygen atoms, respectively. Notably, a shorter distance geometrically corroborates the formation of the 4-coordinated state, as the oxygen atom physically settles into the center of the cation cage. Specifically, in the O3c state, the oxygen atoms are located far from the cation center of mass (large distance) and exhibit a distinct <*111*> orientation. Immediately after transitioning to the O4c state upon polarization switching (Figure 2(d): O3c → O4c), the distance rapidly decreases. This close proximity to the center acts as direct geometric evidence that the 4-coordinated state has formed, initially exhibiting a clear <*100*> orientation (Figure 2(b), (d)). Subsequently, as the simulation progresses under the applied electric field, this <*100*> orientation gradually relaxes into a random orientation while the oxygen maintains its short distance from the origin, as explicitly demonstrated by the continuous time-evolution analysis across the entire switching cycle (**Figure S4**).

Because of this prior relaxation, the O4c oxygen exhibits no specific orientation just before the next polarization switching (Figure 2(c): O4c → O3c). Upon switching, the distance abruptly increases again as the oxygen leaves the center, and the behavior returns from the random state back to the sharp <*111*> orientation characteristic of the O3c state. The perfectly consistent changes across these three microscopic metrics definitively establish that the polarization switching of HZO is not a simple shift like in conventional S:N/S:T models, but rather proceeds with the dynamic mutual exchange of O3c and O4c, accompanied by significant spatial displacements, as its intrinsic driving mechanism.

### 3.3. Microscopic Details of the Polarization Switching Pathway

To determine the specific crystallographic geometric pathways of the oxygen atom exchange behavior confirmed in the previous section, we defined a rectangular cuboid (cation cage) surrounded by 4-coordinated cations and analyzed the displacement trajectories of individual oxygen atoms within the cage (**Figure 3**). The vertical axis represents the y-direction, and the horizontal axis represents the x-direction; the positive and negative displacements in the y-direction are folded and mapped to make the symmetry visually apparent.

As is evident from Figure 3 and **Figure S5** of the Supplementary Information (unfolded plots and y-x, z-x, z-y planar views), this is a result that clearly demonstrates that the behavior corresponding to the so-called "C:N34ex pathway" [16]—which does not cross the cation plane (N), is accompanied by a distinct displacement in the y-direction (C), and involves the mutual exchange of coordination numbers (-34ex)—functions as the actual dynamic process. The oxygen atoms displace while remaining inside the cation cage of the 4-coordinated state without crossing the cation plane.

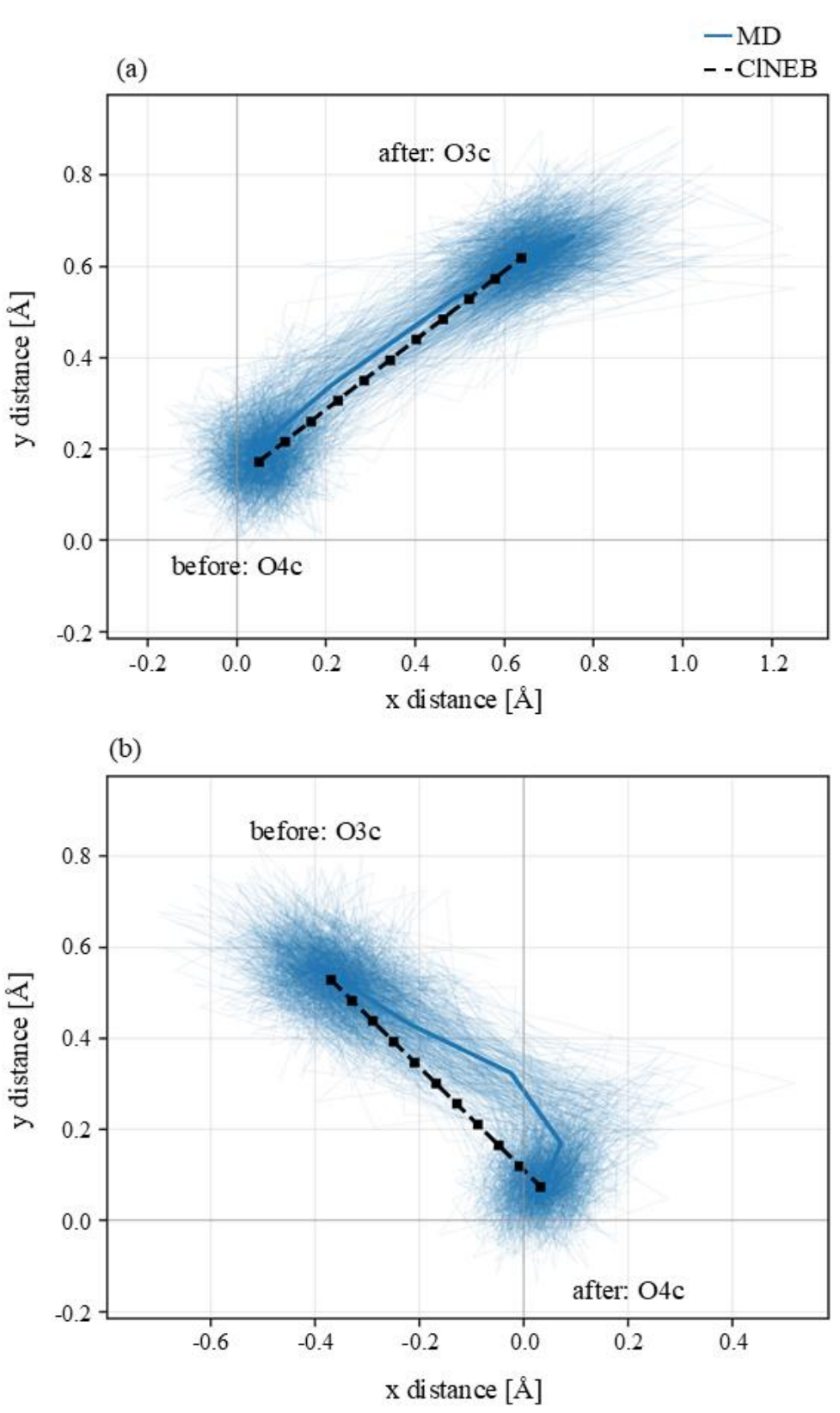

**Figure 3.** Displacement trajectories of individual oxygen atoms within the cation cage during the (a) O4c → O3c process and (b) O3c → O4c process. The trajectories of individual oxygen atoms are shown as thin blue lines, their average pathway (average structure of MD) as a thick blue line, and the trajectories calculated by the CINEB method as black dashed lines.

Observing the trajectories in detail, while the O4c → O3c process in Figure 3(a) traces a relatively linear path, the O3c → O4c process shown in Figure 3(b) reveals that the oxygen atoms do not move in a straight line; rather, they follow a "detour" pathway, first displacing in the negative direction along the x-axis before moving in the positive direction. To eliminate the influence of complex thermal vibrations of individual atoms and verify the intrinsic structural trends, we extracted the average pathway for each process. Specifically, at each MD step, the spatial coordinates of all oxygen atoms undergoing the same transition type (i.e., the O4c → O3c group and the O3c → O4c group) were averaged separately. As a result, the exact same detour trend was confirmed with high reproducibility in the averaged structures as well, ensuring that this is not a transient phenomenon due to thermal fluctuations, but rather an intrinsic transition mechanism unique to this pathway. Therefore, after confirming its validity, this analysis validates the use of the structures at each MD step.

To elucidate the physical origin of this detour behavior, we evaluated the relationship between the cation–oxygen bond lengths and the y-direction distance in both the structures at each MD step and those generated by the ClNEB method [45] (**Figure 4**). The results revealed that the unique detour in the O3c → O4c process is strongly attributed to the formation process of the newly formed fourth cation–oxygen bond. By moving while keeping the distance of this new bond shorter, the oxygen atom is directly influenced—as if being pulled by the newly bonded cation—resulting in this specific curved trajectory.

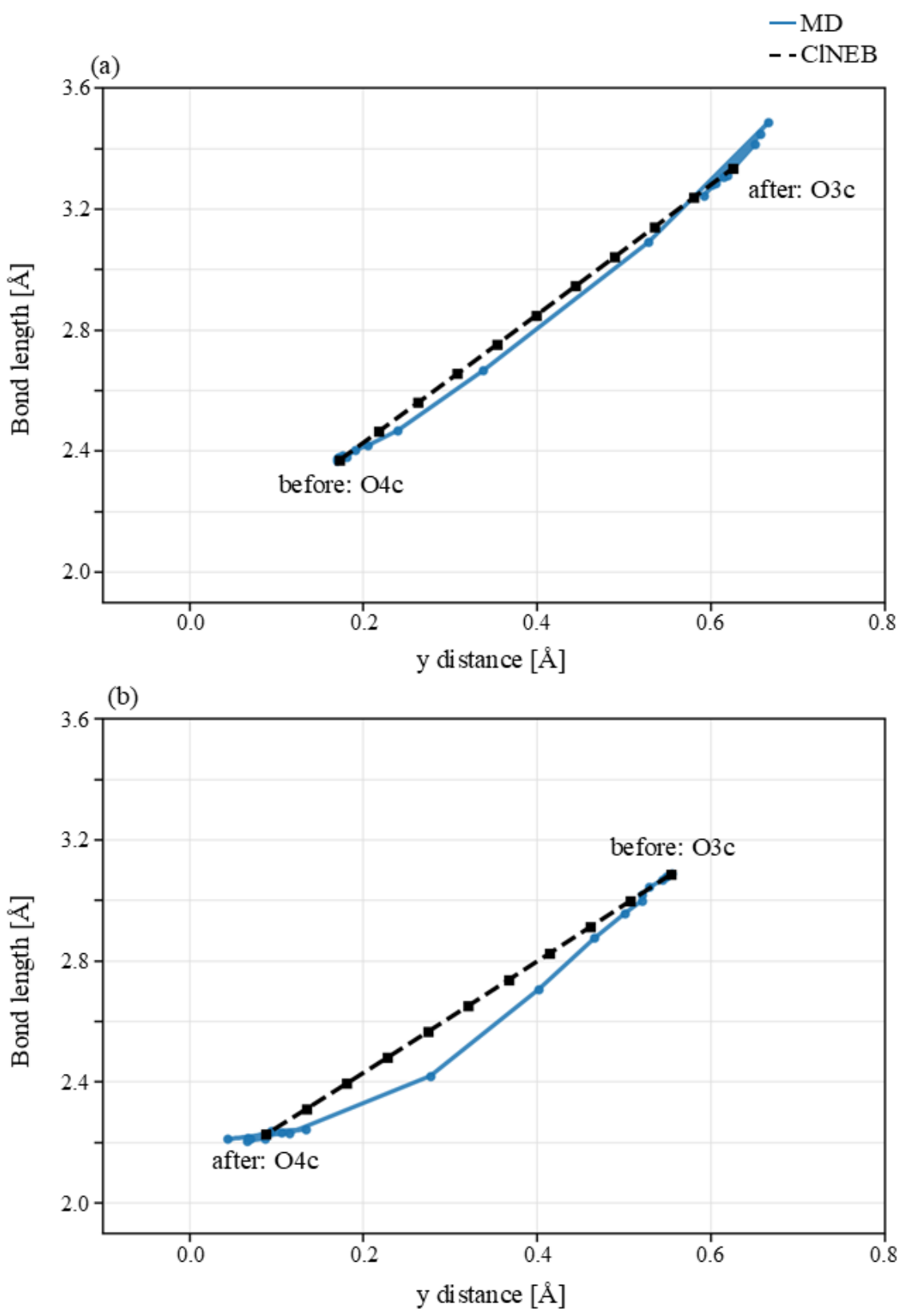

**Figure 4.** Relationship between the cation–oxygen bond length and the y distance of the oxygen atom during the (a) O4c → O3c process and (b) O3c → O4c process. The plots show the specific cation–oxygen pair that exhibits the largest change (accompanied by bond breaking and formation) between the O4c and O3c states in both processes. Structures extracted from the MD simulations are shown in blue, and those generated by the ClNEB method are shown as black dashed lines.

Furthermore, as demonstrated by the analysis in Figure 2 and conceptualized in the state-transition cycle in **Figure 5**, the forward and reverse transition processes are not entirely equivalent time-reversals of each other. This asymmetry is inherently rooted in the distinct orientational and structural characteristics of their respective pathways across the polarization hysteresis loop. Specifically, while the forward O3c → O4c process involves a transition from a *<111>* orientation (State 1) to a *<100>* orientation (State 2), the reverse O4c → O3c process represents a transition from a random orientation (State 3) back to a *<111>* orientation (State 4).

Even when strictly focusing on the *<111>*-oriented O3c states, a prominent spatial difference exists: the initial O3c state before polarization switching (State 1) is located at a relatively short center-of-mass distance ($r_O$) of approximately 0.71 Å, whereas the final O3c state reached after reverse switching (State 4) rests at a significantly longer distance of approximately 0.98 Å. In contrast, during the O4c phase (State 2 and 3), the oxygen atom consistently maintains a close distance of approximately 0.24 Å near the cage center regardless of orientation relaxation. These clear geometric variations indicate that the oxygen atoms experience completely different local environments and geometric constraints during the two switching phases. This structural asymmetry explains why the detour behavior exclusively emerges during the forward O3c → O4c pathway (State 1 → State 2) and is absent in the reverse pathway (State 3 → State 4).

### 3.4. Strain Suppression Mechanism During Polarization Switching

An extremely unique property of HZO is that, despite the drastic microstructural reconfiguration of the oxygen atom coordination numbers (the C:N34ex pathway) occurring within the cell as described above, the change in the macroscopic lattice constant is exceptionally small. **Figure 6** presents the time evolution of the lattice length, unit cell volume, polarization value, and applied electric field against the MD step number. Remarkably, throughout the entire switching cycle under fully relaxed conditions, the dimensional changes remain suppressed to merely around 1–2%. Specifically, during the positive-to-negative polarization switching, the a-axis lattice constant decreases from 5.09 Å to 5.03 Å (a change of −1.04%), while the b- and c-axes contract slightly from 5.29 Å to 5.27 Å (−0.36%) and from 5.04 Å to 5.03 Å (−0.28%), respectively. Consequently, the unit cell volume decreases by 2.27 $Å^3$, moving from 135.59 $Å^3$ to 133.32 $Å^3$ (−1.67%). Similarly, during the negative-to-positive switching, the a-, b-, and c-axes decrease from 5.09 Å to 5.03 Å (−1.21%), from 5.29 Å to 5.27 Å (−0.30%), and from 5.04 Å to 5.03 Å (−0.22%), respectively, accompanied by a 2.33 $Å^3$ volume reduction from 135.60 $Å^3$ to 133.27 $Å^3$ (−1.72%). These quantitative evaluations clearly demonstrate that the minor volume contraction (1.67–1.72%) accompanying polarization switching is predominantly governed by the roughly 1% contraction along the a-axis, while the deformations along the b- and c-axes are virtually negligible (less than 0.36%). This intrinsic geometric stability—where the lattice dimensions perpendicular to the primary switching axis undergo minimal distortion even without external constraints—strongly suggests that the C:N34ex switching mechanism can proceed smoothly even under severe epitaxial constraints. To verify this hypothesis, we next investigate the polarization switching dynamics under in-plane clamped conditions, simulating the actual mechanical environment of thin-film devices.

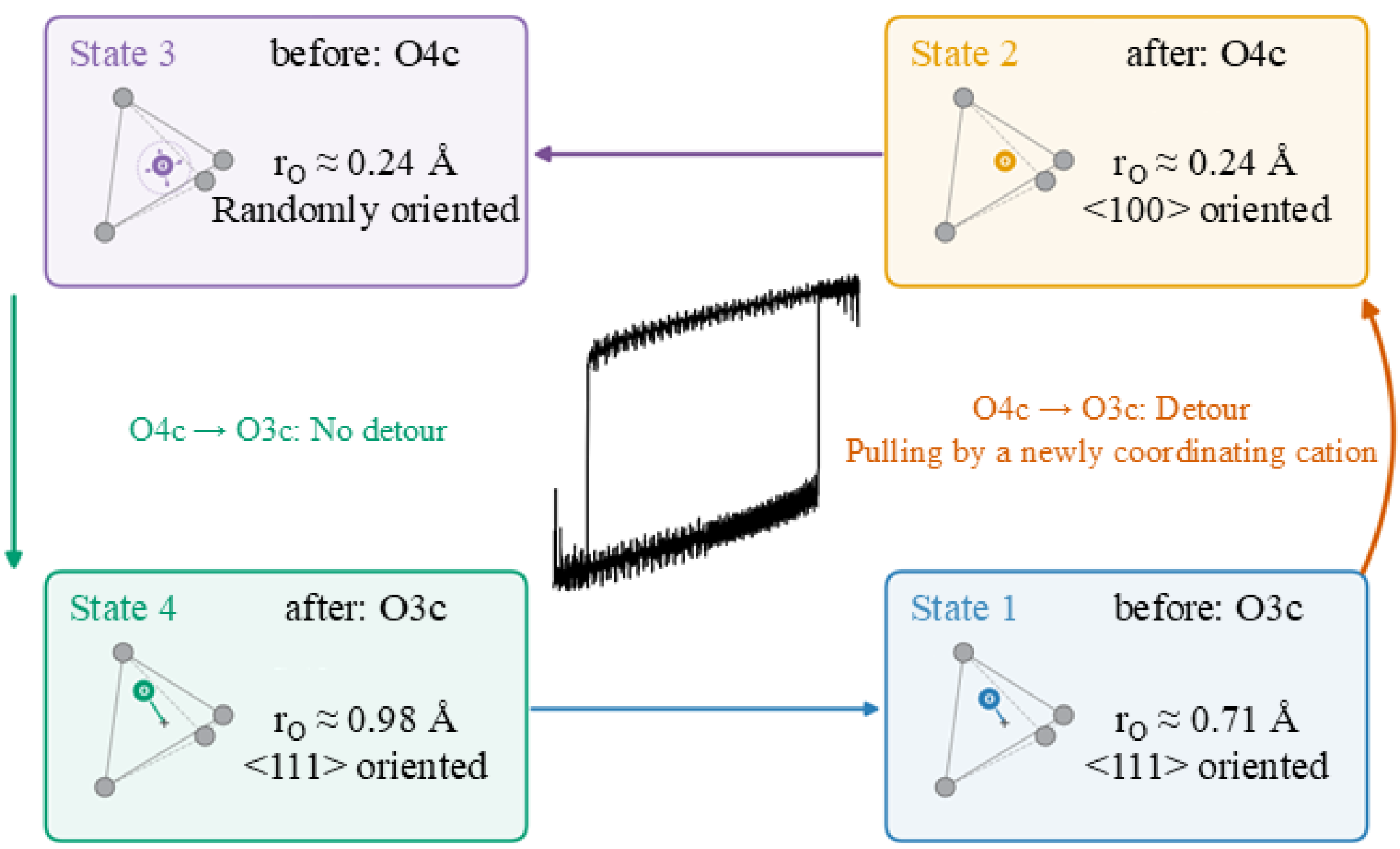

**Figure 5.** Schematic illustration of the state-transition cycle for oxygen atoms during polarization switching. The central *P–E* hysteresis loop is correlated with the four distinct structural and orientational states (State 1 to State 4) of the oxygen atom within the cation cage across the switching cycle. State 1 and State 4 correspond to the O3c states exhibiting <111> orientation, while State 2 and State 3 represent the O4c states with <100> and random orientations, respectively. The colored arrows indicate the transition pathways along the hysteresis loop: the forward transition accompanied by the detour trajectory (State 1 → State 2, orange arrow) and the reverse transition without detour (State 3 → State 4, green arrow).

To verify the correlation between this lattice flexibility and ferroelectric properties, we calculated the *P-E* hysteresis loop under the condition where the b and c-axis is completely clamped (fixed) (**Figure S6**). The $E_c$ of HZO with the b and c-axis fixed is 8.18 MV/cm, showing almost no change compared to the fully relaxed state (7.59 MV/cm). The calculation result showing that excellent ferroelectric properties can be stably maintained even under macroscopic shape constraints from the outside is highly consistent with the experimental fact that HZO does not lose its ferroelectricity even in nanometer-scale ultra-thin films or under strong clamping effects from dissimilar material interfaces. This is an important finding that microscopically explains its excellent retention characteristics (ultra-thin film stability).

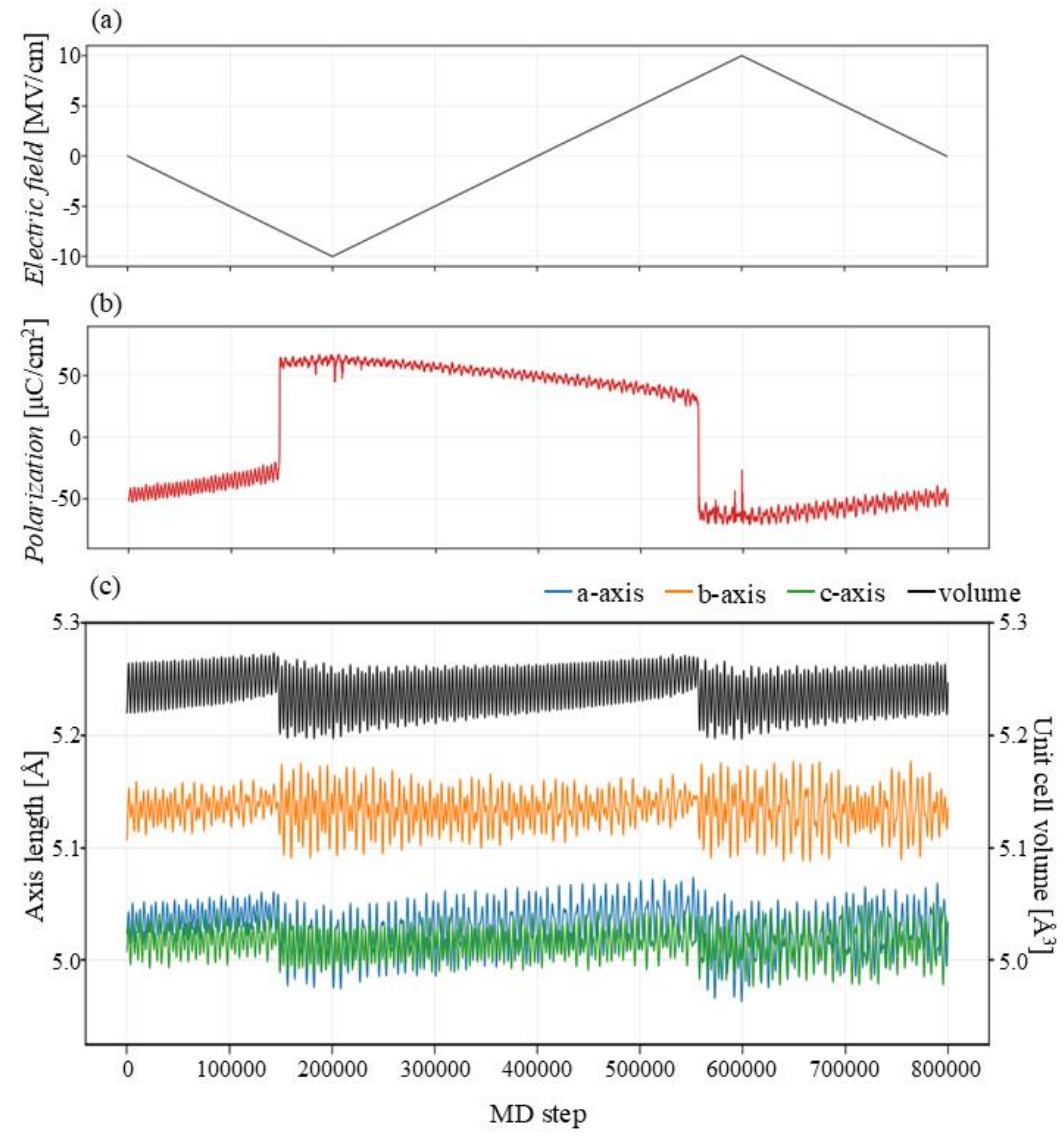

**Figure 6.** Three-tier plot showing the evolution of various macroscopic indicators against the number of MD steps during electric-field-induced MD simulations: (a) applied electric field, (b) spontaneous polarization, and (c) lattice lengths (a-, b-, and c-axis lengths) along with the percentage change in unit cell volume. In (c), the a-, b-, and c-axis lengths are shown on the left y-axis in blue, orange, and green, respectively, while the percentage change in the unit cell volume is displayed on the right y-axis (in black).

To uncover the microscopic factors giving rise to this outstanding clamping resistance, we focused on the dynamic volume changes of the local tetrahedra formed by the 4-coordinated cations. **Figure 7** shows the progression of the tetrahedral volume against MD steps (Figure 7(a): O4c → O3c, Figure 7(b): O3c → O4c). The analysis revealed that the local volumetric "expansion" that occurs along with the O4c → O3c transition is perfectly received and offset within the cell by the volumetric "contraction" caused by the simultaneously progressing O3c

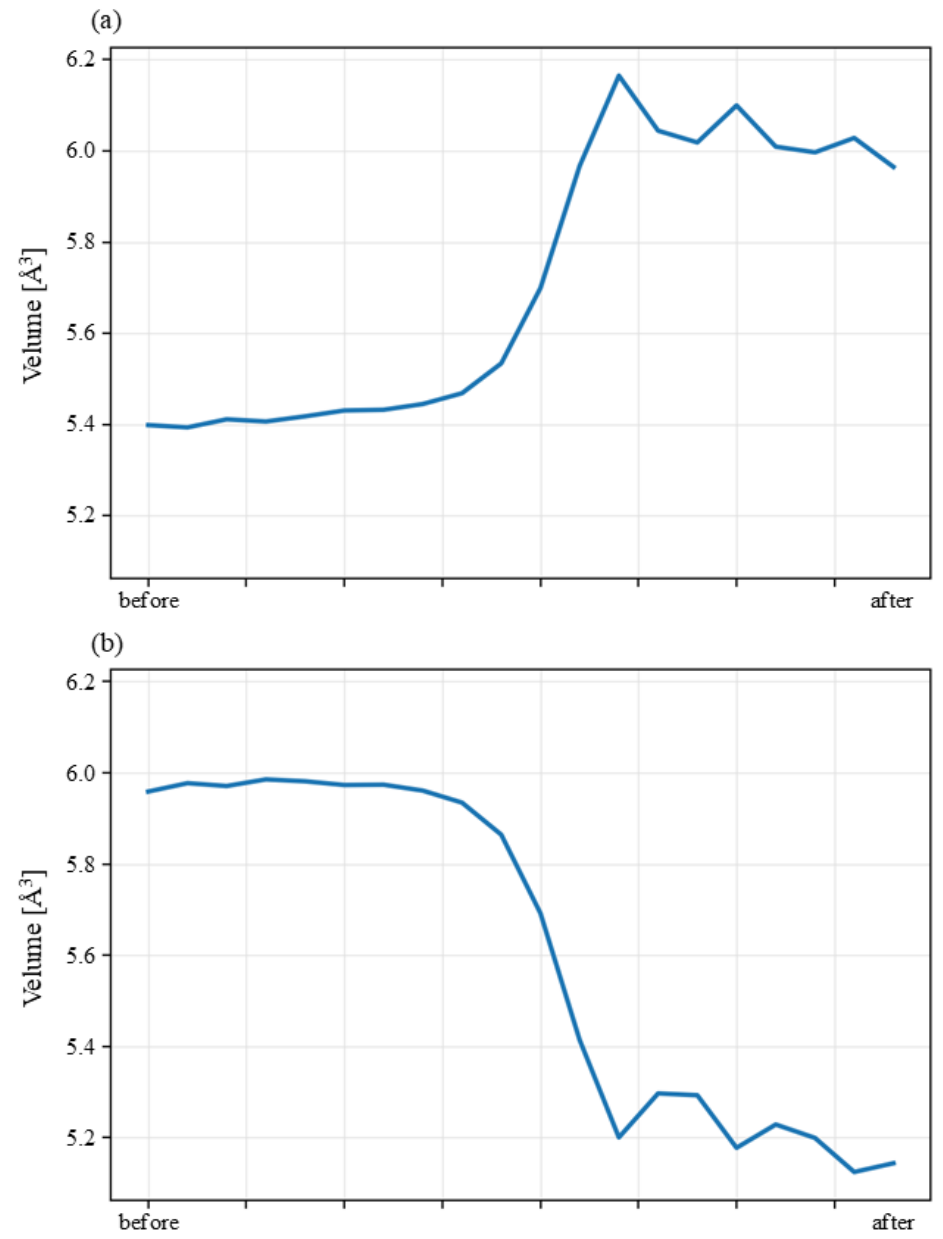

**Figure 7.** Evolution of the tetrahedral volume formed by the 4-coordinated cations against MD steps during the (a) O4c → O3c transition process and (b) O3c → O4c transition process. The data are plotted for a total of 20 structures before and after polarization switching.

→ O4c transition. In other words, during the polarization switching of HZO, oxygen atoms with different coordination environments are exchanged cooperatively and complementarily within the same cell. Because this "internal self-compensation mechanism of local volume change" is at work, polarization switching becomes possible without altering the macroscopic lattice constant at all. The unique internal compensation mechanism discovered in this study is considered to be one of the fundamental reasons why HZO can stably retain ferroelectric properties that are ideal for device applications, even in the form of ultra-thin films.

## 4. Conclusion

In this study, we constructed electric-field-coupled MD simulations utilizing a machine learning interatomic potential, thereby elucidating the detailed microscopic dynamics of the polarization switching mechanism in HZO ferroelectrics, which had previously remained unresolved. The developed MACEField model reproduces the energy, interatomic forces, and BECs from first-principles calculations with exceptionally high accuracy. Furthermore, it precisely captures the lattice-dynamical phonon characteristics, successfully describing a physically reasonable *P-E* hysteresis loop that correctly reflects the intrinsic properties of a defect-free ideal lattice. The key insights obtained from our simulations are summarized as follows:

1. Validation of the Polarization Switching Mechanism Involving Coordination Number Mutual Exchange
   We demonstrated that, unlike conventional simple atomic shift models (the S:N/S:T models), polarization switching in HZO proceeds with the dynamic mutual exchange of O3c and O4c within the same cell as its intrinsic driving mechanism. Coupled analyses of displacement vectors, pole figures, and degrees of orientation clearly captured a series of dynamics wherein the oxygen orientation dynamically transitions during the switching cycle from the <*111*> orientation (initial state) to the <*100*> orientation (immediately after the first switching), then to a random orientation (precursor process of the second switching), and finally back to the <*111*> orientation (after the second switching).
2. Identification of Transient Detour Behavior within the C:N34ex Pathway
   Crystallographic geometric pathway analysis dynamically demonstrated for the first time that migrating oxygen atoms displace while remaining inside the 4-coordinated cation cage without crossing the cation plane, following the so-called "C:N34ex pathway." Particularly during the O3c → O4c transition process, we discovered a unique trajectory where the oxygen atoms do not move linearly but rather take a temporary "detour" in the x-direction. We determined that this detour behavior is a consequence of the strong influence of transient bond formation, specifically reflecting the tendency to maintain a shorter distance for the newly forming fourth cation-oxygen bond.
3. Elucidation of the Physical Origin of Strain-Free Polarization Switching and Ultra-Thin Film Stability
   Throughout the entire polarization switching process, HZO exhibits an outstanding clamping resistance that distinguishes it from conventional ferroelectrics, as its macroscopic lattice constants remain nearly constant. Investigating the origin of this macroscopic strain-free property from a local structural perspective revealed that an "internal self-compensation mechanism of local volume changes" is at work; the local volumetric "expansion" of the tetrahedra occurring during the O4c → O3c process is complementarily offset within the cell by the simultaneous "contraction" of the O3c → O4c process.

These findings provide a clear physical answer from the perspective of microscopic crystal dynamics to a major question in device applications: why HZO stably retains excellent ferroelectric properties even in nanometer-scale ultra-thin film states or under strong clamping constraints from interfaces. The dynamic analysis approach and the uncovered microscopic mechanism presented in this work are expected to serve as crucial guidelines for the material design of next-generation ultra-high-density, low-power ferroelectric memories and nanoelectronic devices.

**Acknowledgements**

This study was supported by the Ministry of Education, Culture, Sports, Science and Technology (MEXT, 26K22608, 26K01205, 25K24645), and New Energy and Industrial Technology Development Organization (NEDO). P.-Y.C. would like to acknowledge the support of JST SPRING (Grant Number JPMJSP2108). The computation was carried out using the computer resource offered by Research Institute for Information Technology, Kyushu University.

**Author Contributions**

R.S. and P.-Y.C. conceived the idea and designed the research. R.S. performed the simulations, analyzed the data, and prepared the figures. R.S. wrote the original draft of the manuscript. T.M. supervised the project and acquired funding. All authors discussed the results, reviewed, and edited the manuscript.

**Data Availability Statement**

The trained machine-learning force field models, Born effective charge models, and simulation trajectories generated in this study are available from the corresponding author upon reasonable request.

# Supporting Information

**Supporting Information for: Transient Detour and Cooperative Oxygen Exchange in the Polarization Switching of Ferroelectric $Hf_{0.5}Zr_{0.5}O_2$**

[1]Ryotaro Sahashi, [1]Po-Yen Chen, and [1,2]Teruyasu Mizoguchi
[1]Department of Materials Engineering, The University of Tokyo, Tokyo, Japan
[2]Institute of Industrial Science, The University of Tokyo, Tokyo, Japan

This Supporting Information provides detailed validation results of the machine-learning force-field (MLFF) and Born effective charge (BEC) models, as well as representative structural configurations and supplementary visualization of the electric-field-induced molecular dynamics (MD) simulations.

## S1. Validation of the MACEField Model

To investigate the ferroelectric switching behavior of $Hf_{0.5}Zr_{0.5}O_2$ at the atomic scale, we fine-tuned a MACEField model that describes the potential energy surface (PES) by learning total energies, interatomic force, and BECs.

### *S1.1 Prediction Accuracy of Forces, Energies, and BECs*

To quantitatively evaluate the predictive accuracy of the MACEField model constructed in this study, the correlations of total energy, interatomic forces, and BEC were assessed against a test dataset based on VASP calculations (**Figure S1**). Each predicted value exhibited extremely high correlation, with Mean Absolute Errors (MAE) for Force, Energy, and BEC being sufficiently low at 12.1 meV/Å, 25.8 meV, and $7.33 \times 10^{-3}$ e, respectively, confirming an exceptionally high level of reproducibility.

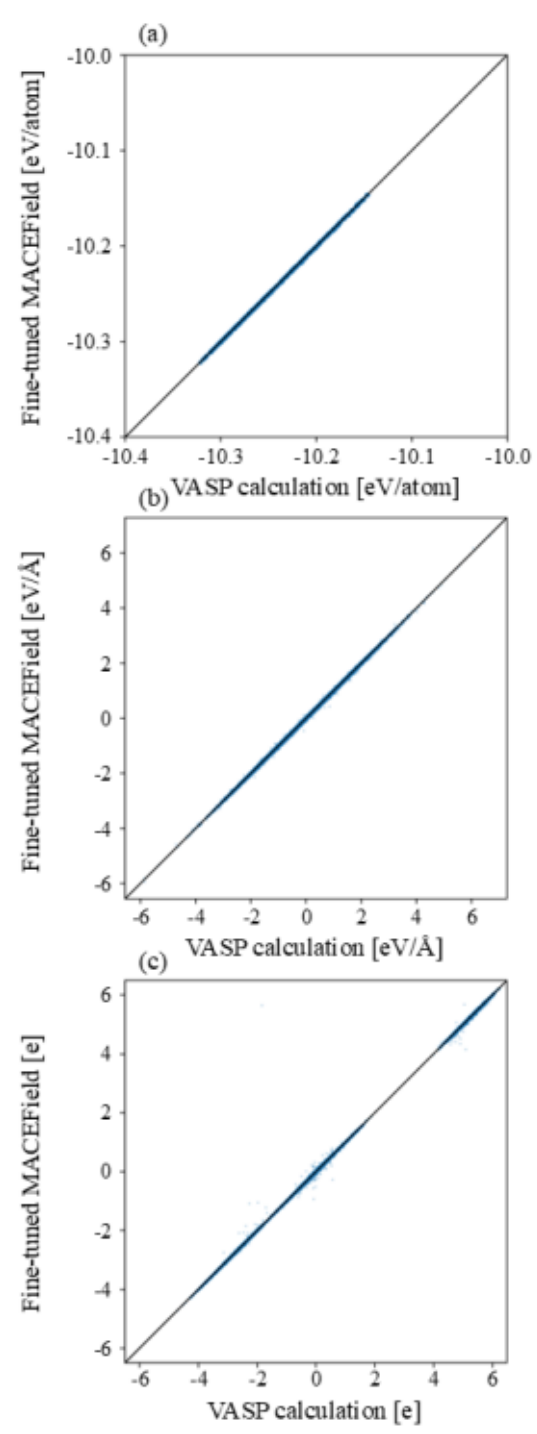

**Figure S1.** Correlation plots of (a) energy, (b) atomic forces, and (c) BECs for the fine-tuned MACEField model. The horizontal axis of each graph represents the VASP calculation results, and the vertical axis represents the predicted values by MACEField.

### *S1.2 Validation of Phonon Dispersion and Lattice Dynamics in Each Crystalline Phase*

To ensure the structural stability and lattice dynamical validity of the trained force field, phonon dispersion relations were calculated for the major crystalline phases of HZO ($Pca2_1$, $P2_1/c$, $P4_2/nmc$, and Pbca) (**Figure S2**). The predicted phonon band structures showed excellent agreement with the VASP calculation results across the entire Brillouin zone, with MAEs for each phase being 0.039, 0.037, 0.039, and 0.028 THz, respectively. This demonstrates that the model possesses outstanding predictive accuracy from a lattice dynamics perspective as well.

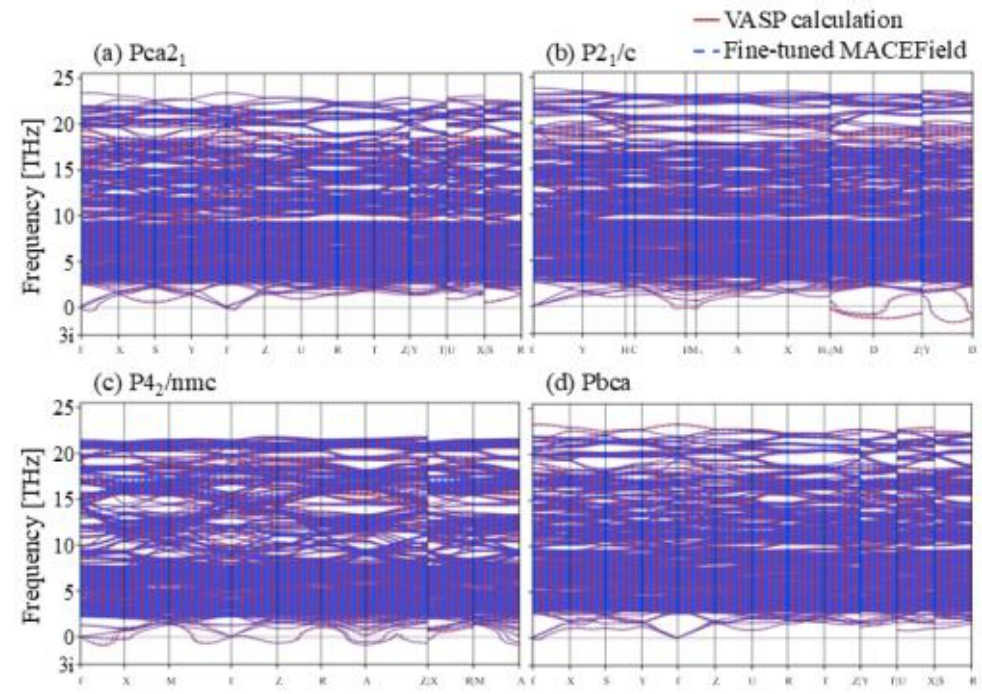

**Figure S2.** Comparison of phonon dispersion relations for the four primary crystalline phases of HZO ($Pca2_1$, $P2_1/c$, $P4_2/nmc$, and Pbca) calculated using the MACEField model and VASP calculation results.

*S1.3 Temperature Dependence of Coercive Field and Relative Permittivity*

To further validate the reliability of the fine-tuned MACEField model under finite-temperature and electric-field-applied MD conditions, we evaluated the macroscopic temperature-dependent electromechanical responses of the ferroelectric $Pca2_1$ phase. Specifically, MD simulations were performed across various temperatures to determine the temperature dependence of the coercive field ($E_c$) and the relative dielectric constant ($\varepsilon_r$), as shown in **Figure S3**.

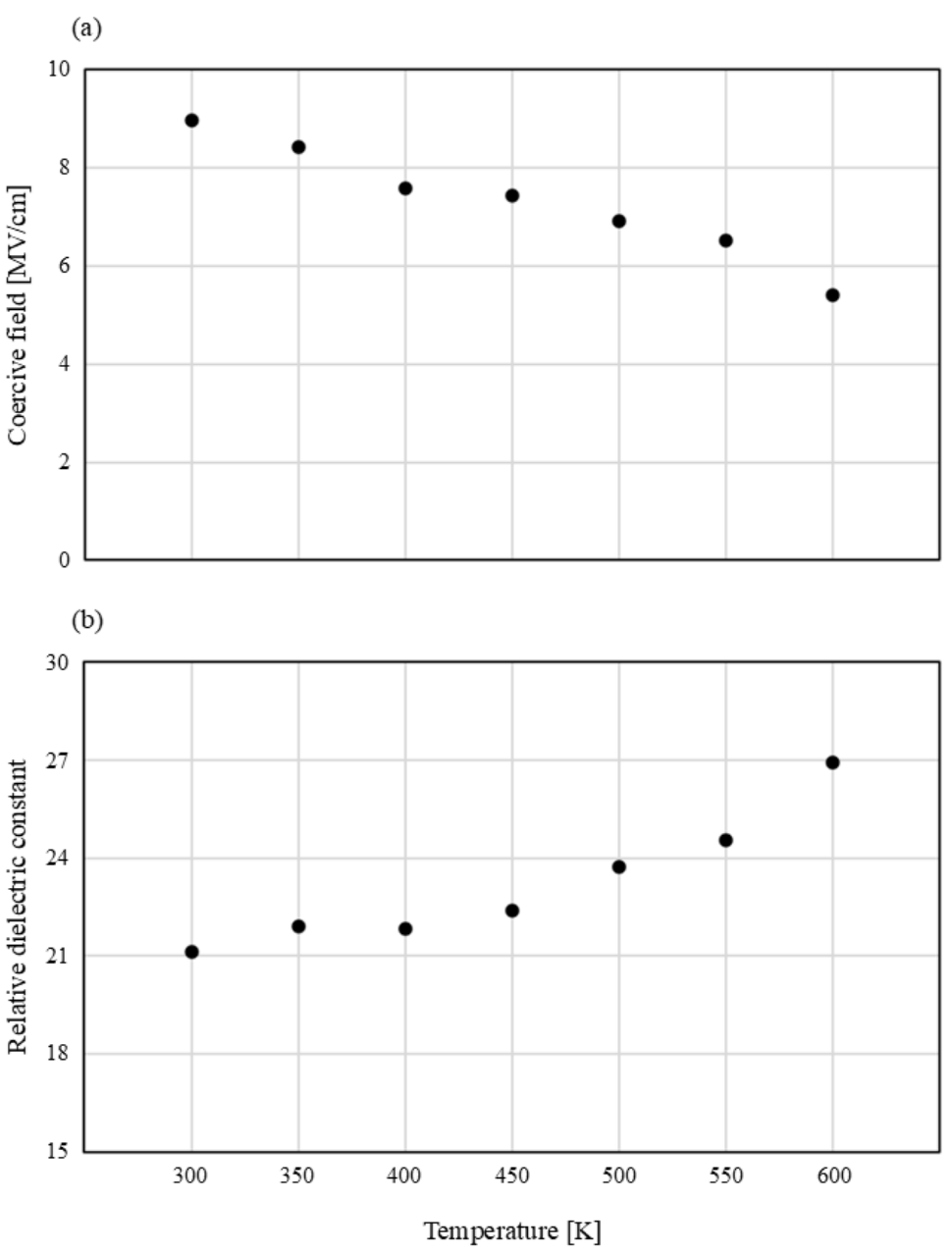

**Figure S3.** Temperature dependence of the macroscopic electromechanical properties of the HZO $Pca2_1$ phase calculated using the fine-tuned MACEField model.
(a) Coercive field $E_c$) and (b) relative dielectric constant ($\varepsilon_r$) plotted as a function of temperature.

As the temperature increases, the calculated coercive field monotonically decreases (Figure S3(a)), reflecting the thermally assisted reduction of the energy barrier required for polarization switching. Concurrently, the relative permittivity exhibits a continuous upward trend with increasing temperature (Figure S3(b)). These simulated macroscopic behaviors are in excellent qualitative and quantitative agreement with experimental measurements reported for HZO thin films. [1,2] This confirms that our MACEField model accurately captures not only 0 K ground-state properties and lattice dynamics but also finite-temperature thermodynamic and dielectric responses under applied electric fields, establishing a highly robust foundation for the subsequent switching simulations.

## S2. Dynamics and Geometric Trajectories of Oxygen Atom Displacement Accompanying Polarization Switching

*S2.1 Continuous Evolution of the Degree of Orientation Throughout the Polarization Switching Cycle*

To confirm the dynamic displacement behavior of oxygen atoms over a longer time scale, we analyzed the degree of orientation throughout the entire MD simulation (**Figure**

**S4**). Tracking a series of polarization switching cycles clearly captured the transient evolution of the oxygen atom orientation as the switching progressed. Specifically, a continuous dynamic sequence was confirmed (*<111>* → *<100>* → random → *<111>*): the initial *<111>*-oriented state temporarily shifted to a *<100>* orientation, subsequently relaxed into a random orientation under the applied electric field, and finally recovered its original *<111>* orientation upon the completion of the switching.

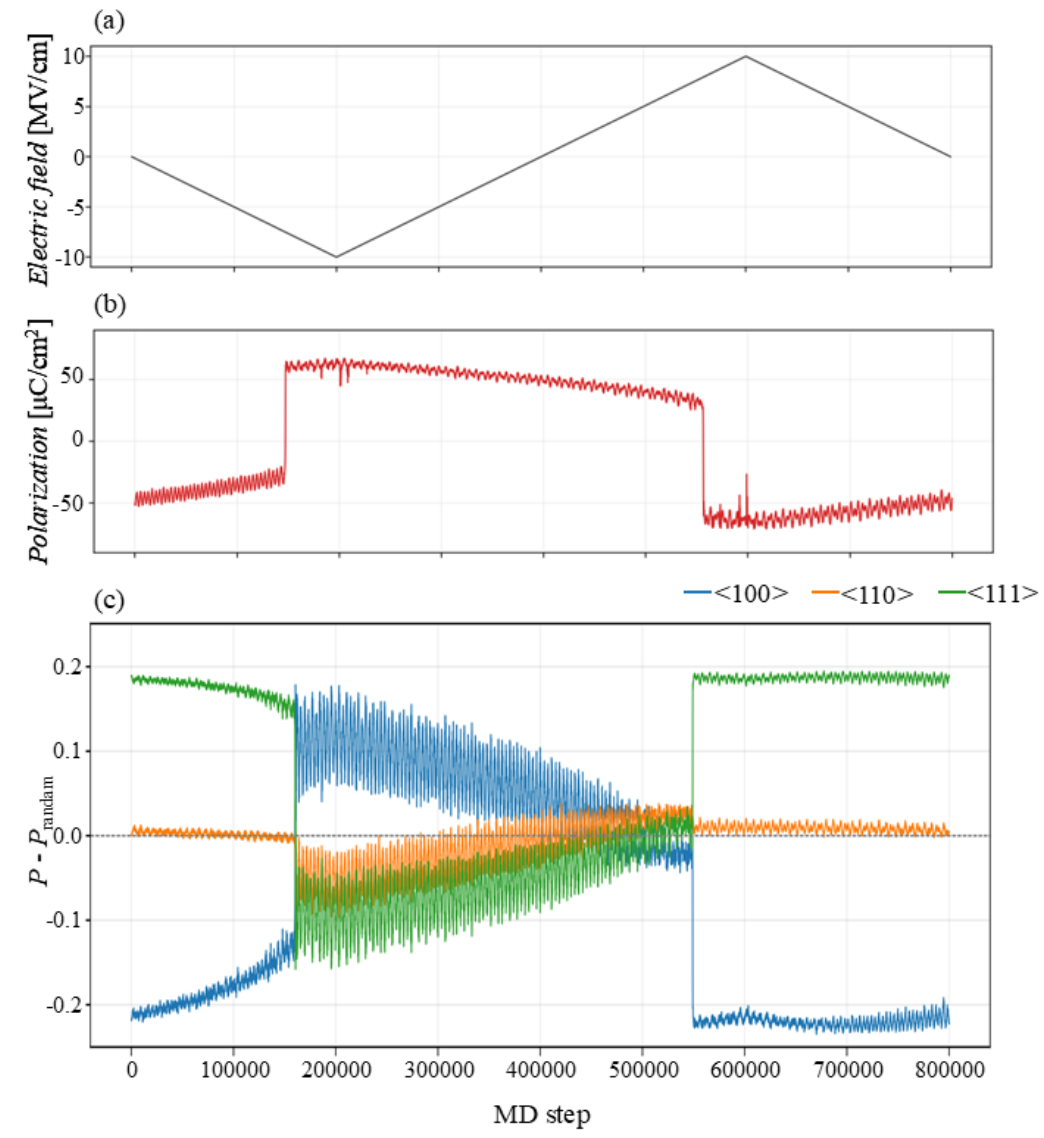

**Figure S4.** Dynamic evolution of various indicators during the polarization switching cycle. Plots of (a) spontaneous polarization, (b) degree of orientation $P_{\langle uvw \rangle}$, and (c) degree of orientation minus the random baseline, covering the range from the first polarization switching to the second polarization switching. The horizontal axis of all graphs is the number of MD steps (STEP), and each direction in the orientation-related plots (c) is indicated by *<100>* (blue), *<110>* (orange), and *<111>* (green).

### *S2.2 Geometric Projections of Oxygen Trajectories and the C:N34ex Mechanism*

To visualize the spatial movement of oxygen atoms along the polarization switching pathway in detail, displacement trajectories projected onto the z-x, z-y, and y-x planes were plotted (**Figure S5**). In the two-dimensional projections from the z-x and z-y planes, the trajectories of the oxygen atoms superficially appear to intersect. However, for an oxygen atom to actually pass through the plane formed by the cations, a displacement of approximately 1.26 Å from the center of mass in the x-direction is required. Since the movement distance in the observed trajectories is insufficient to cross this plane, it can be determined that the oxygen atoms do not pass through the cation plane. This geometric feature strongly corroborates that the polarization switching is not a simple atom-through-plane phenomenon, but rather proceeds via a coordination number 3 and 4 mutual exchange (C:N34ex) mechanism within the cation cage.

Additionally, focusing on the projection from the y-x plane, it can also be observed that the oxygen atoms follow a unique "detour" pathway rather than a straight line during the O3c → O4c process.

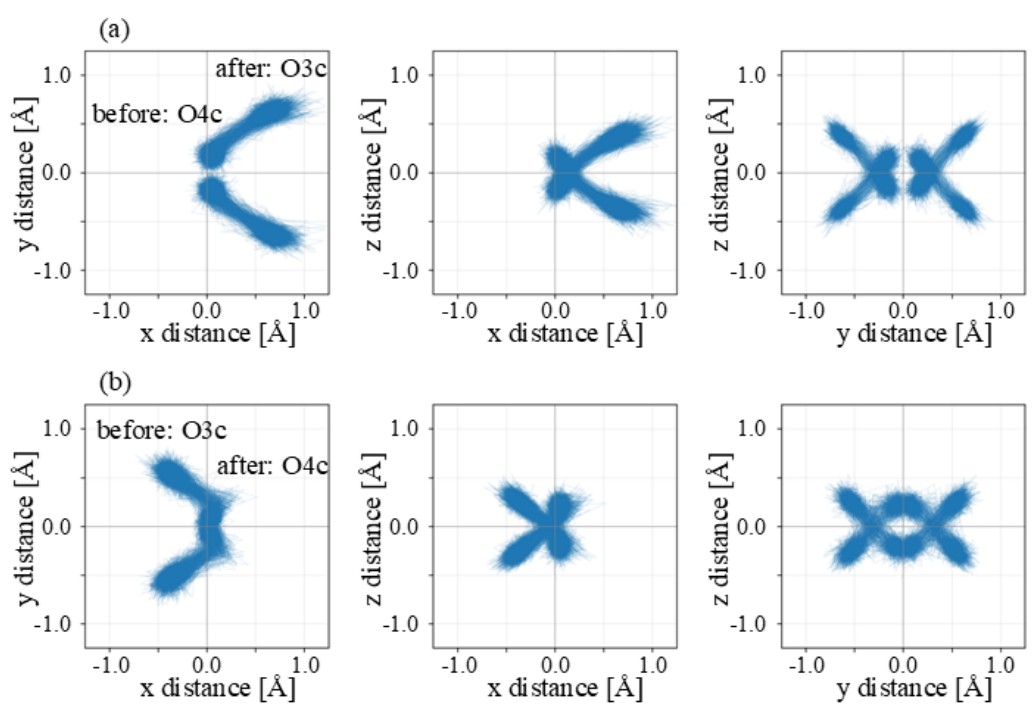

**Figure S5.** Projections of oxygen atom displacements onto each plane (y-x, z-x, and z-y planes) during the (a) O4c → O3c process and (b) O3c → O4c process. The origin of the oxygen atom displacement is defined as the center of mass of the four surrounding cations when the targeted oxygen atom is in the O4c state.

## S3. Ferroelectric Response Under Geometric Constraints (Fixed bc-Axes)

To verify the constraint effects typically imposed by substrates in thin-film states, electric-field-induced MD simulations were performed under conditions where the cell dimensions of the b- and c-axes were fixed (clamped) (**Figure S6**). Compared to the fully relaxed state (unconstrained), a distinct *P-E* hysteresis loop is maintained, and the magnitude of the coercive field remains almost unchanged. Furthermore, while the clamped b- and c-axis lengths are held constant, the unconstrained a-axis exhibits oscillatory behavior in response to the applied electric field; however, no macroscopically large dimensional changes occur. These results demonstrate that the polarization switching process in HZO is highly robust even under geometric constraint conditions.

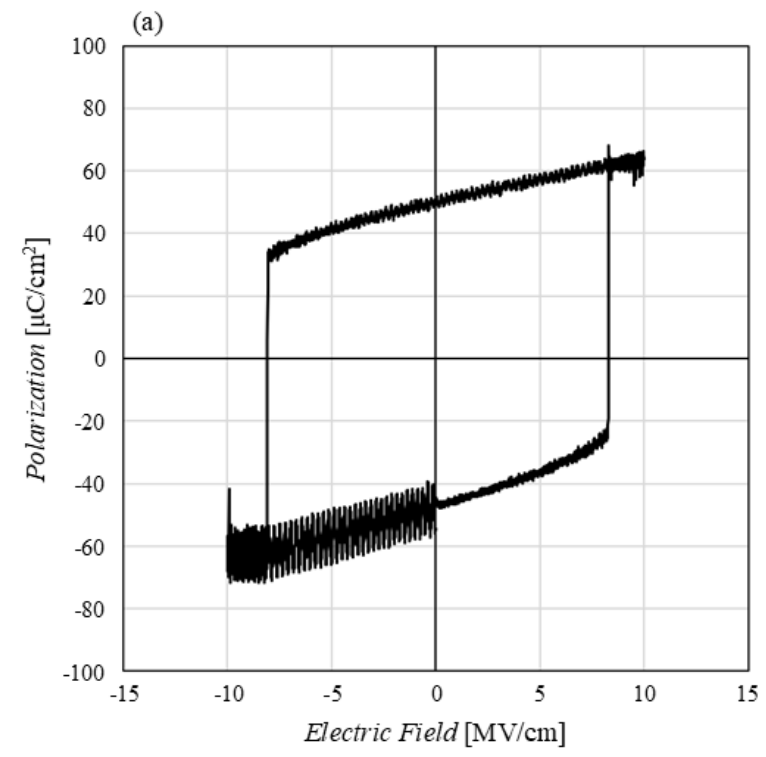

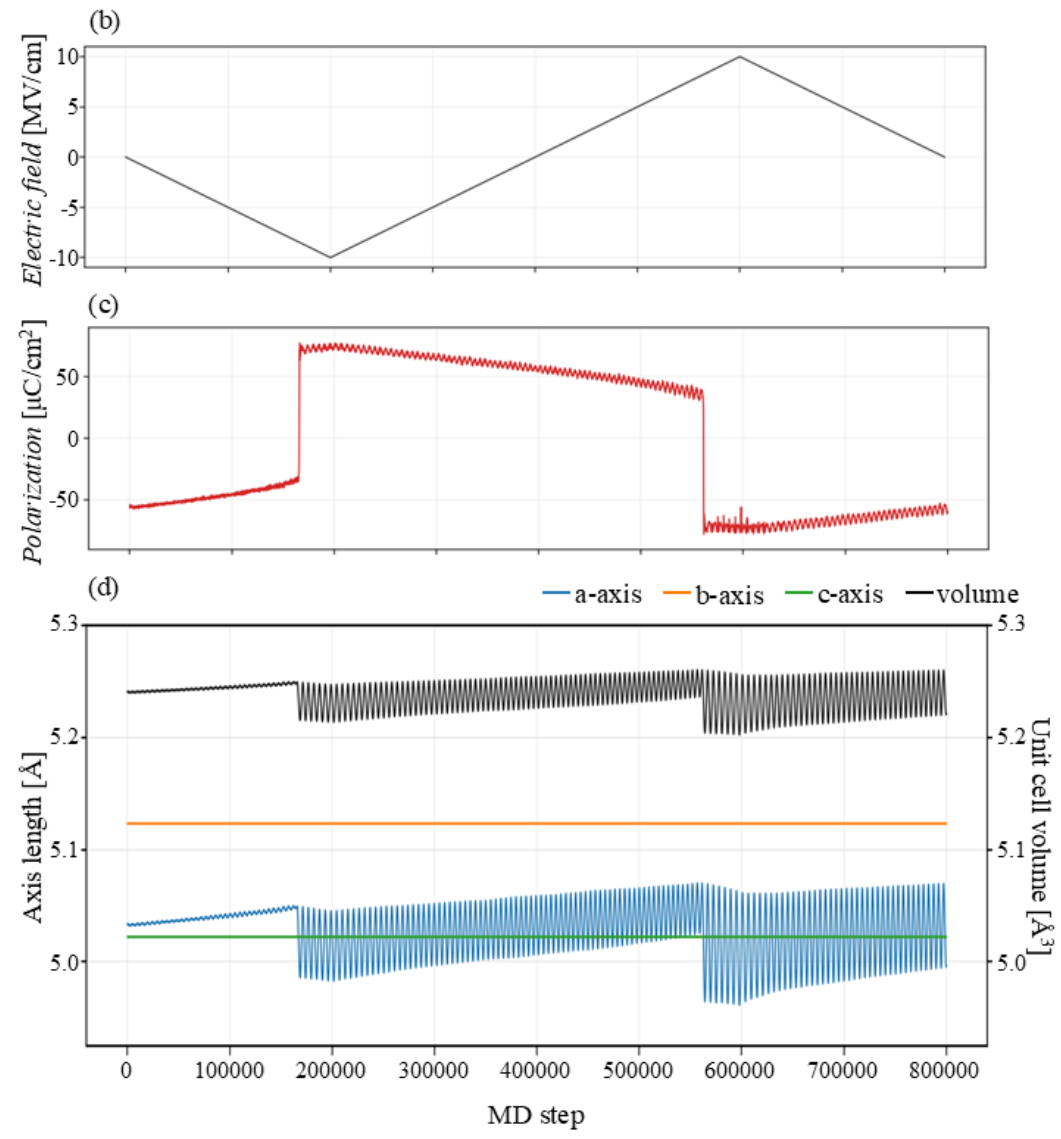

**Figure S6.** Various plots under geometrically constrained conditions (bc-axis clamped state): (a) *P-E* hysteresis loop under the bc-axis clamped state, and the three-tier plot showing the evolution of (b) applied electric field, (c) spontaneous polarization, and (d) lattice lengths (a-, b-, and c-axis lengths) along with the percentage change in unit cell volume. In (d), the a-, b-, and c-axis lengths are shown on the left y-axis in blue, orange, and green, respectively, while the percentage change in the unit cell volume is displayed on the right y-axis (in black).

## S4. Evaluation of Supercell Size Dependence

To determine the optimal system size necessary to appropriately describe the macroscopic ferroelectric response without size artifacts in the electric-field-induced MD simulations, comparative calculations were performed using $Pca2_1$ supercells consisting of 96, 768, and 2,592 atoms (**Figure S7**).

First, comparing the *P-E* hysteresis loops calculated for each cell size (Figure S7(a)), the 96-atom model exhibits highly unstable behavior with large fluctuations in the polarization value in regions where a large electric field is applied. However, as the number of atoms increases, this instability is significantly suppressed, leading to smooth switching behavior.

Next, plotting the relationship between the supercell size and the coercive field (Figure S7(b)) reveals that the coercive fields for each cell size transitioned to 5.53, 7.55, and 7.59 MV/cm, respectively. The coercive field increases compared to the small-scale 96-atom cell and clearly converges to a stable, size-independent value at 2,592 atoms. This result validates that the 2,592-atom supercell used for the primary analyses in this study is sufficiently large to stably describe the polarization switching dynamics.

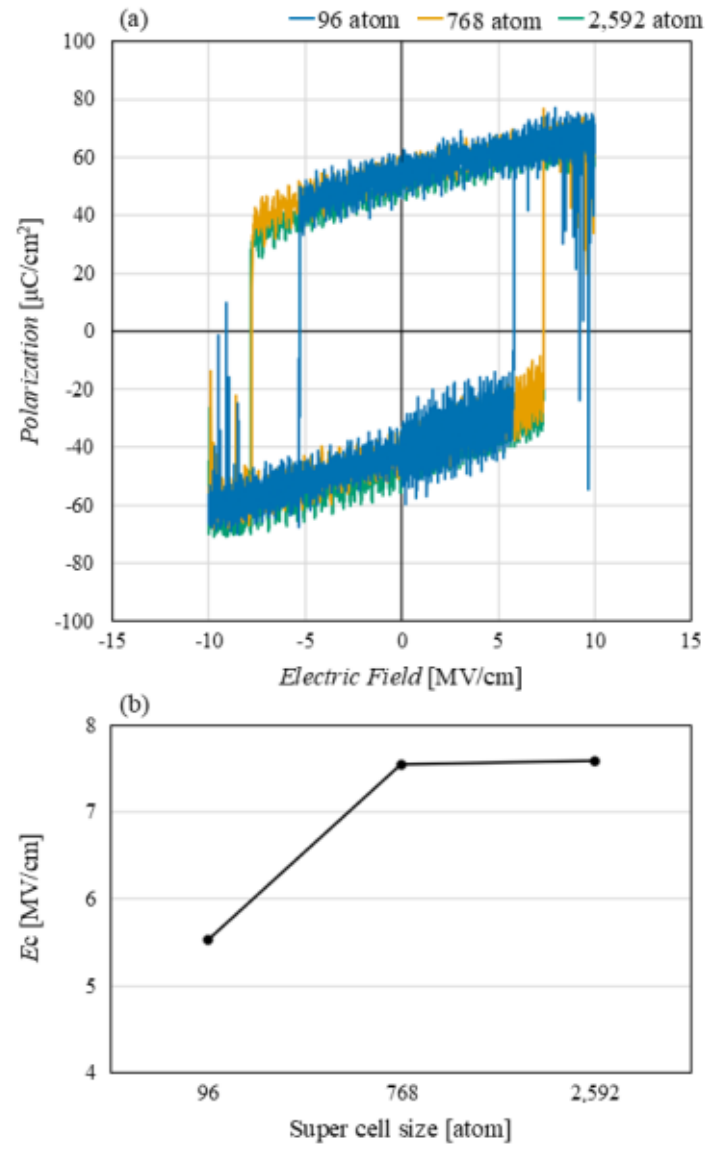

**Figure S7.** Supercell size dependence of the ferroelectric properties in the electric-field-induced MD simulations. (a) *P-E* hysteresis loops calculated for $Pca2_1$ supercells containing 96, 768, and 2,592 atoms. (b) Plot showing the relationship between the supercell size (horizontal axis) and the $E_c$ (vertical axis).